\RequirePackage{times}
\RequirePackage{fullpage}
\RequirePackage{ifthen}

\documentclass[onecolumn,11pt]{article} 

\renewenvironment{abstract}{%
	\setlength{\parindent}{0in}%
	\setlength{\parskip}{0in}%
	\bfseries%
}{\par\vspace{-6pt}}

\newcommand{\spacing}[1]{\renewcommand{\baselinestretch}{#1}\large\normalsize}
\spacing{1}

\usepackage{titlesec}
\usepackage[a4paper, total={7.5in, 10in}]{geometry}
\usepackage{bm}
\usepackage[latin1]{inputenc}
\usepackage{amsmath}
\usepackage{amsfonts}
\usepackage{amssymb}
\usepackage{graphicx}
\usepackage[usenames,dvipsnames]{xcolor}
\usepackage[caption=false]{subfig}
\usepackage{notoccite}
\usepackage[numbers,sort&compress]{natbib}
\usepackage[caption=false]{subfig}

\titleformat*{\section}{\bfseries}
\titleformat*{\subsection}{\bfseries}

\begin{document}
	
	\begin{flushleft}
		{\Large			
			\textbf\newline{ \bf \textsf{Multimorbidity as a multistage disease process} }
		}
		\newline
		\\
		{\large Anthony J. Webster}
		\\
		\vspace{0.3cm}
		{\sl \large
			Department of Statistics, 
			24-29 St Giles', 
			University of Oxford, 
			Oxford, OX1 3LB,
			UK.}
		\\
		\vspace{0.1cm}
		{\sl Email}: anthony.webster@stats.ox.ac.uk
	\end{flushleft}

\begin{abstract}
There is a growing proportion of people with several disease conditions 
(``multimorbidity''), placing increasing demands on healthcare systems. 
One hypothesis is that clusters of diseases may 
arise from shared underlying disease processes (shared ``pathogenesis''), 
whereby the presence of one disease indicates the state of disease 
progression to several related disease types. 
This article explains how this hypothesis can be tested using observational 
data for disease incidence.
Specifically, a multistage model is used to 
test whether two diseases can have a ``shared stage'' or ``step'', 
before either disease can occur, and how the unobserved rate of 
this step can be determined.  
The approach offers a simple method for studying multiple 
diseases and identifying shared underlying causes 
of multiple conditions, 
and is illustrated with published data and numerical examples. 
The fundamental mathematical model is analysed to compare key 
statistical properties such as the expectation and variance 
with those of independent diseases. 
The main results do not need an understanding 
of the underlying mathematics and can be appreciated by a non-expert. 
\end{abstract}
\vspace{0.25cm}
\noindent
{\bf {\textit{Significance:}} 
It is widely believed that there are shared underlying pathways that 
can lead to several disease types (shared ``pathogenesis''), and this 
may explain observed clusters of disease types. 
This article shows how this hypothesis can be tested for a pair or 
cluster of diseases, using observational data of disease incidence.}
Longitudinal studies of cohorts such as UK Biobank \cite{Bycroft2018}, 
usually consider the risk of a single disease in individuals who are otherwise 
disease-free, in the sense of not having any conditions that would be expected to 
modify the individuals' risk of the disease being studied. 
However there is an increasingly recognised need to be able to authoritatively 
study and understand disease-risks in individuals with two or more conditions 
(``multimorbidity'' \cite{Skou2022}). 
Here we consider the statistical properties of disease incidence when two or 
more diseases share a process or step that is necessary for either of 
them to occur. 
The result is a simple graphical test using age-dependent incidence data, 
to determine whether the shared-step model is consistent with the observed 
rates of two or more diseases.  
The approach allows quantitative estimates for the rates of the different 
steps in the ``multistage'' disease process, and has the potential to 
provide a simple explanation for why we would expect to see some 
individuals with clusters of several disease types.
An example that is discussed in Section \ref{Diab}, considers type 2 diabetes 
and atherosclerotic heart disease (AHD) risk. 
In this example, a combination of exposures and aging first lead to insulin resistance and 
hyperglycemia. 
This intermediate stage where insulin resistance and hyperglycemia develop, is also 
associated with endothelial damage (damage to the cells lining blood vessel walls), 
lipid deposition, and inflammatory processes that are necessary for 
atherosclerotic heart disease (e.g. see \cite{Poznyak2020,Ormazabal2018,Jellinger2007}). 
As a result, once type 2 diabetes has been diagnosed, several necessary steps 
needed for AHD will already have occurred, and the observational 
data are consistent with this (figure \ref{pnasfig}, Section \ref{Diab}).
Conceptually, the model considers two or more diseases that have a shared 
set of disease processes (a shared ``step'' or ``stage''), that is necessary for 
all the diseases' development, but may not necessarily be observed or observable. 
The diseases can only occur after the first shared 
step has occurred, and the subsequent steps can occur with different rates and risk 
factors. 
From a mathematical perspective, each ``stage'' or ``step'' could in principle 
involve several intermediate steps or processes, the result of which has a 
cumulative distribution function for its time to occur. 
This ``multi-stage'' model \cite{Webster2022} was originally motivated by the 
observation that disease-risks in people with one or more diseases were increased 
by a very similar factor of between $1$ and $2$ \cite{Webster2023}, and the finding 
that multistage disease processes could be very common \cite{Webster2022}. 
Multistage models such as that studied here \cite{Webster2019}, 
have previously been used to model the age-dependent incidence of many 
diseases \cite{Gerovska2020, Webster2022, Gerovska2022, Elgarresta2024}, with dedicated studies of 
diseases that 
include Cancer \cite{Nordling1953, AD1954, Knudson1971, Luebeck2002, Michor2006, Devarajan2024}, 
Autoimmune disease \cite{Goodnow2007},
Amyotrophic Lateral Sclerosis (ALS) \cite{Chio2018, Corcia2018, Vucic2019, Licher2019, Vucic2020, Garton2021, Ziser2024}, 
Parkinson's \cite{LeHeron2021}, 
Alzheimer's \cite{Licher2019},
Renal disease \cite{Obeid2022,Guran2024},
Aneurysmal haemorrhage \cite{Ruigrok2024},
and COVID \cite{Pearce2021}. 
There has also been considerable recent work on ``multi-state'' models 
that explicitly model the transitions between each 
step \cite{Kalbfleisch1988,Gentleman1994,Jackson2003,Therneau2024,Broomfield2024}, 
using Markov-chain transitions between each stage. 
A representative selection of work includes studies on 
ALS \cite{Gil2007}, 
Hepatitis C \cite{Sweeting2010}, 
Parkinson's \cite{Geman2013},
Alzheimer's \cite{Brookmeyer2019,Zhang2019}, 
Cardiovascular disease \cite{Chen2023,Urbut2024},
and the Liver \cite{Yu2024}. 
An important distinction between this approach and the work presented here, is that 
several distinct and potentially rate-limiting stages are  
considered here as single stages. 
This is beneficial for simplifying the underlying concept 
of the disease processes, and for subsequent modelling. 
The conceptual model, although underpinned by detailed mathematics, 
is intended to be understandable by clinicians and epidemiologists. 
A reader who wishes to grasp the key points but omit the mathematical detail, can 
read the introduction, figure \ref{DAG}, sections \ref{Diab} and \ref{Detection}, 
and the conclusions in section \ref{Discussion}. 
The next section \ref{Diab} starts by discussing the example of diabetes and cardiovascular 
disease, using previously analysed and published data \cite{Webster2022}. 
Section \ref{Detection} then describes how similar analyses can be used 
to detect evidence for a shared step, given data for the incidence of 
two diseases, and explains how to interpret the resulting plots. 
The comparatively simple procedure involved, is the main result of the paper, 
offering a simple but powerful technique to use observational data to detect evidence 
for diseases that share a step that is common to both diseases' pathogenesis. 
The main mathematical properties of the model are summarised in 
section \ref{Model}, and derived in appendix \ref{gf}. 
The properties of the model are explored further for the specific example of 
processes with constant hazard rates in Section \ref{examples}, and a 
second example that explores the relationship with conventional multistage 
models is in appendix \ref{ex2}.  

\section{Diabetes and cardiovascular disease}\label{Diab}

\begin{figure}
	\centering
	\includegraphics[width=0.9\linewidth]{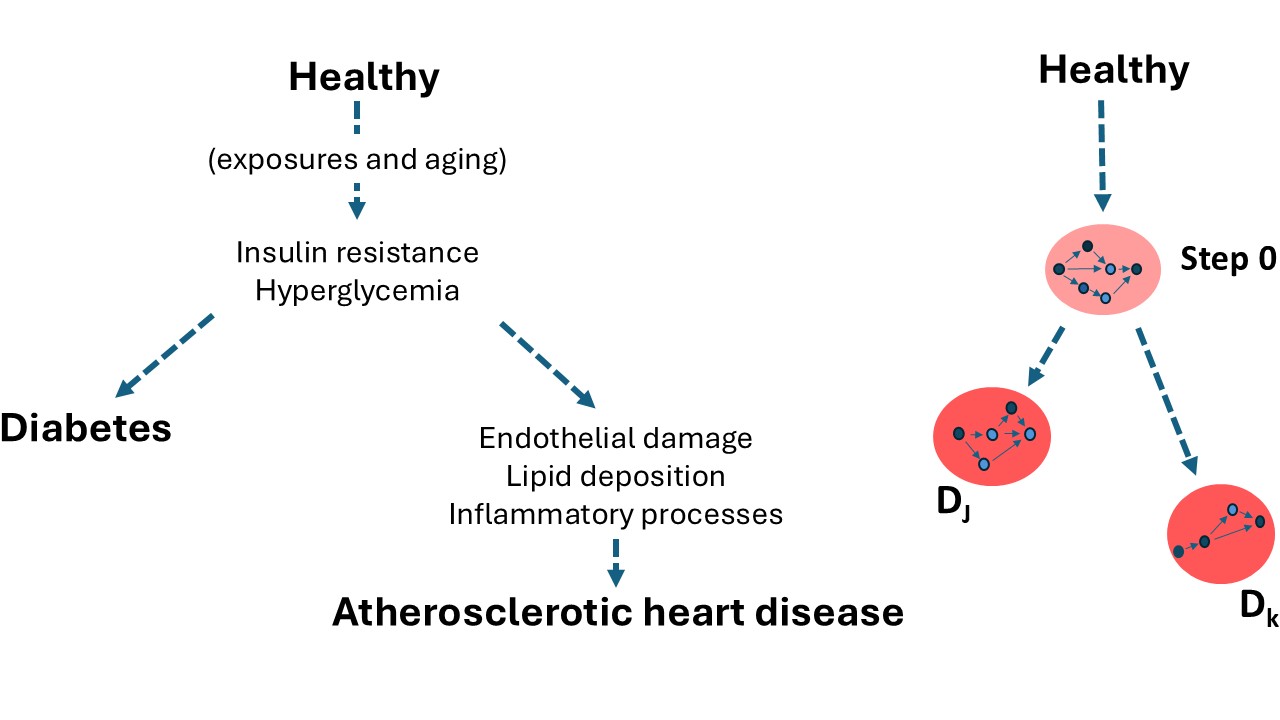}
	\caption{(Left) Diabetes arises from insulin resistance and 
		hyperglycemia, that also  
		cause damage that is necessary for Atherosclerotic Heart Disease (AHD). 
		(Right) 
		More generally, we will consider a model for diseases that have a 
		shared ``stage'' or ``step'', that is necessary for all the diseases 
		to occur (insulin resistance and hyperglycemia in the example here), 
		but is not necessarily observed or observable. 
		Each step may consist of 
		several distinct and potentially rate-limiting stages.  
	}\label{DAG}
\end{figure}

\begin{figure}[h]
	\begin{center}
		\includegraphics[width=.48\linewidth]{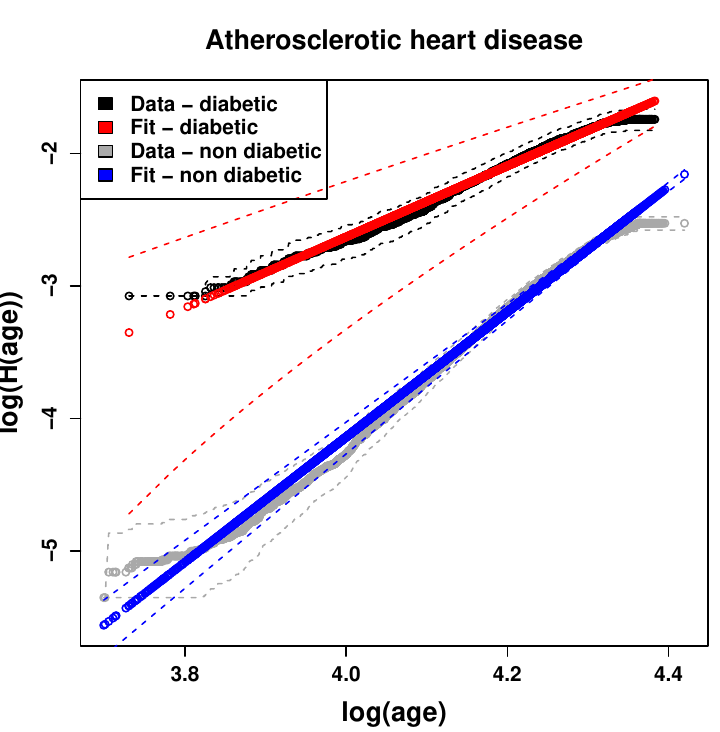}
		\includegraphics[width=.48\linewidth]{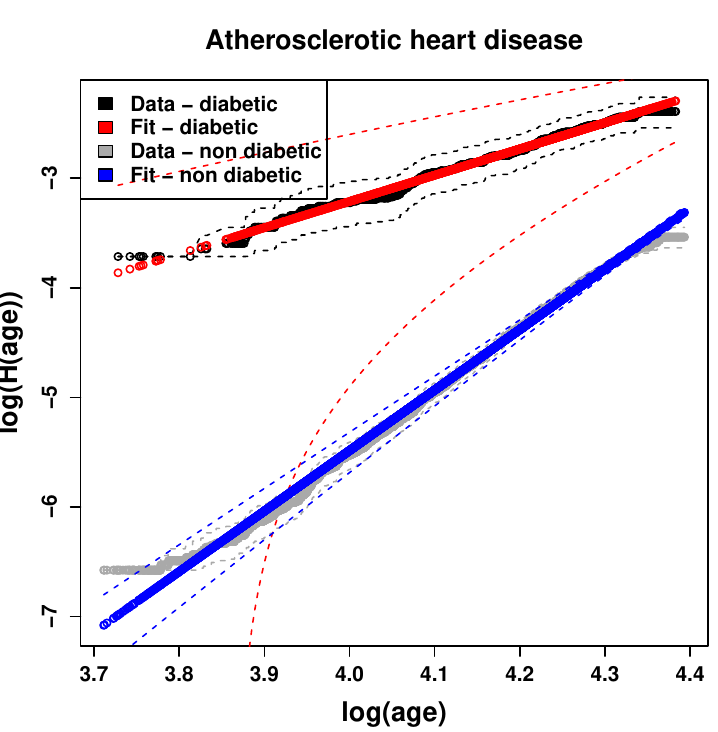}
	\end{center}
	\caption{
		The log of cumulative hazard function $log(H(age))$, are plotted versus $log(age)$ for 
		atherosclerotic heart disease (AHD) in men (left), and women (right), with data stratified 
		by diabetic status (most cases of which will be type 2). 
		The vertical shift combined with a reduction in slope seen for diabetics, 
		is what we would expect if both diabetes and AHD share a necessary step or 
		process prior to their onset \cite{Webster2022}.
		(Reproduced from the Supplementary Material in Webster and Clarke \cite{Webster2022})
	}\label{pnasfig}
\end{figure}

Consider the example of diabetes and atherosclerotic heart disease (AHD), 
for which published data already exist \cite{Webster2022}. 
AHD involves several necessary disease processes \cite{Poznyak2020,Ormazabal2018,Jellinger2007}, 
broadly summarised as:
\begin{enumerate}
	\item Endothelial dysfunction - through inflammation and damage to the 
	endothelial cells lining blood vessels. 
	\item Lipid accumulation - within the arterial walls and via macrophage foam cells. 
	\item Inflammation - with immune cells infiltrating and activating in the arterial walls. 
	\item Plaque formation and instability.
	\item Disease incidence - thrombosis, ischemia, embolisation. 
\end{enumerate} 
Some of these processes can occur concurrently with each other, but others 
only become important after necessary earlier processes  such as 
lipid accumulation and inflammation have sufficiently developed. 
As a result, AHD disease development can be regarded as a multi-stage 
disease process, with the ``rate-limiting'' steps being the slowest 
disease processes that determine the observed rate of disease 
incidence \cite{Webster2019,Webster2022}. 
Diabetes onset shares several disease processes that are also necessary 
for AHD. 
In particular, the insulin resistance and hyperglycemia that lead to 
type II diabetes are also associated with endothelial damage, 
lipid deposition, and inflammation. 
As a result, by the time type II diabetes is detected, early stages 
of AHD are likely to have begun (figure \ref{DAG}). 
Support for this picture is provided by figure S4 of the Supplementary Material 
in reference \cite{Webster2022}, part of which is reproduced here (fig. \ref{pnasfig}). 
The incidence rates for AHD in figure \ref{pnasfig} are above those of non-diabetics, 
and show a reduced slope with respect to age. 
This is what we would expect to see if the diagnosis of diabetes can be regarded as 
a proxy for exposures that have increased the rate of early-stage AHD processes 
sufficiently that they are no longer rate-limiting; i.e. sufficient damage has 
already occurred that the early steps in the AHD process are not influencing 
the rate of disease onset. 
A simplified causal diagram of the assumed steps in the disease process is shown 
in figure \ref{DAG}. 
In principle the observed change in AHD disease rates in those diagnosed with 
diabetes could be due to diabetes-medication or another resulting change in 
lifestyle. 
However, if e.g. medication caused a proportional change in the hazard rate of AHD, 
then the slope would be displaced vertically, but it would not cause a change in slope.  
The observed combination of an increase in rate and reduction in slope, 
would need the processes leading to diabetes to increase the disease rates and the medication to lead to a reduced slope of the plot. 
This possibility seems unlikely, but cannot be ruled out, 
and illustrates why a change in slope is {\sl necessary} for the presence 
of a shared step but {\sl is not sufficient} to prove its existence. 

\section{Data analysis: Detecting and quantifying a shared step}\label{Detection}

A large proportion of diseases can be studied as multistage disease 
processes \cite{Webster2022}, with incidence rates that can be modelled 
with a Weibull distribution and approximated as a power of age.
Using a multistage disease model as an example, this section describes 
the differences in observational data of disease incidence to expect if 
diseases do, or do not, share a step in their development as in 
figure \ref{DAG}. 

\subsection{Qualitative picture}

If two diseases have incidence rates that are consistent with a multi-stage disease 
process (with incidence rates $\sim t^q$, for an age $t$ and power $q$), then 
plotting the log of incidence rate versus the log of 
age $t$ will give a straight line (provided any censoring and 
left-truncation of the data are suitably accounted 
for \cite{Webster2022}). 
For such diseases $j$ and $k$, a simple test for whether they share a first step is to:
\begin{enumerate} 
	\item[(i)]{Determine incidence rates for disease-free individuals,  
		with a survival analysis 
		and a plot of the log of cumulative hazard 
		versus log of age (e.g. see \cite{Webster2022}, and 
		the right plot in figure \ref{hazFig}).}  
	\item[(ii)]{Repeat the procedure for  
		incidence rates of disease $j$, but now solely for 
		individuals where disease $k$ has 
		already occurred (or vice versa for $k$ given $j$ has occurred). 
		(Left plot in figure \ref{hazFig}.)} 
\end{enumerate}
For the latter situation it is essential to correctly account 
for left-truncation \cite{Webster2022}. 
If the diseases share a first step, then this plot will have a different slope than it did when disease incidence is plotted for 
individuals without prior disease. 
If the plots are simply displaced vertically up or down with respect to those for 
disease-free individuals, then this would indicate that the occurrence of disease $j$ 
has modified the risk of $k$, but if the slope is unchanged then they would     
{\sl not} share a step. 
More generally, consider diseases $j$ and $k$ that do not share a 
first step and need not involve multistage diseases processes, but 
the presence of either disease modifies the incidence rates (e.g. 
due to changed exposures), in a way 
that can be modelled with proportional hazards. 
Because it has been found that many commonly studied exposures can be modelled 
with a proportional hazards framework, this will often be a reasonable modelling 
assumption. 
If the diseases {\sl do not} share a first step then the baseline hazard 
for incidence rates of disease $j$ in disease-free individuals, 
and for disease $j$ after disease $k$ has been observed, will 
be the same (with the proportional hazards adjustment 
accounting for changes in exposures due to disease $k$). 
This may allow conventional proportional hazards methodologies 
to be used to test whether the incidence of diseases $j$ and $k$ 
are consistent with the assumption of the diseases being independent of each 
other (not sharing a step), by testing whether the baseline 
hazard functions for the two cases are consistent with the 
hypothesis of being the same. 
The following will focus on the more specific model where 
each step in the process can be approximated as a power of 
time, as is the case for traditional multistage models of disease. 

\subsection{Quantitative picture}

\begin{figure*}[h]
	\centering 
	\includegraphics[width=0.98\linewidth]{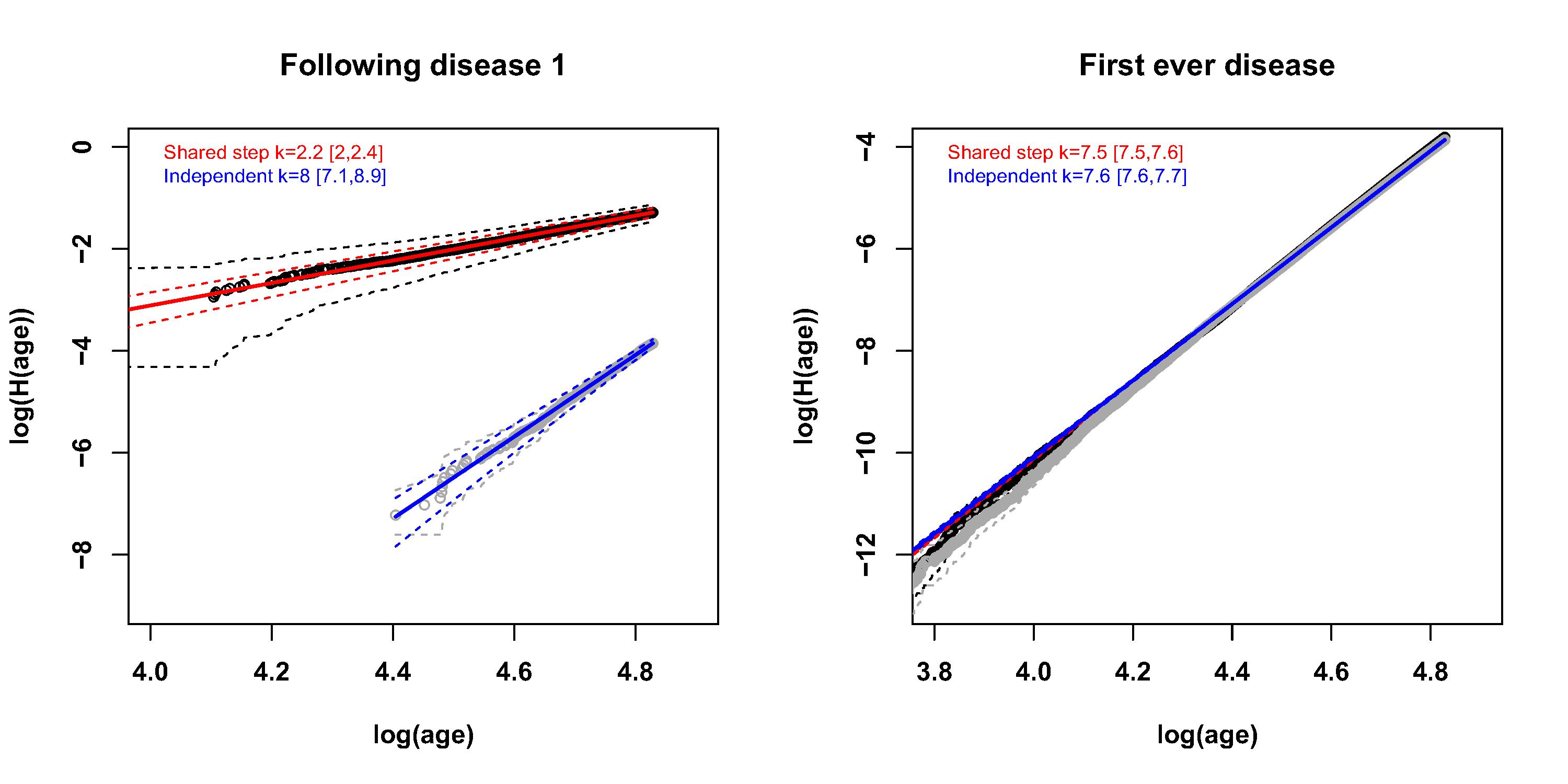}
	\caption{
		For the first observed disease 
		(right figure), incidence rates for 
		independent diseases and diseases with a shared 
		step are equivalent.  
		After one of the diseases have occurred (left figure), 
		then disease risks are the same if the diseases are independent (red), 
		but higher and characterised by a different slope if the 
		diseases share a step (blue). 
	}\label{hazFig}
\end{figure*}

More specifically, consider a multistage disease process with a cumulative 
distribution function (cdf) for the first stage that can be approximated 
as $F_0(t)\simeq \lambda_0 t^{q_0 +1 }$, and 
cdfs for subsequent stages $F_{0j}(t)\simeq \lambda_j t^{q_j +1}$. 
The approximation is reasonable because disease risk for 
many diseases can be approximated by this simple ``multistage'' model \cite{Webster2022}. 
Then Appendix \ref{rates} shows that the cdf for disease $j$ in a previously 
disease-free individual , is, 
\begin{equation}
F_j(t)=\lambda_0\lambda_j c_j(q_0,q_j) t^{q_0+q_j+2}
\end{equation}
where $c_j(q_0,q_j)$ is independent of age or the rate parameters 
$\lambda_0$ and $\lambda_j$.
Similarly for disease $k$, 
\begin{equation}
F_k(t)=\lambda_0\lambda_k c_k(q_0,q_k) t^{q_0+q_k+2}
\end{equation}
Both of these can be estimated from observational data by considering the 
age at the first disease onset (in the conventional epidemiological sense, that 
subsequent data would be censored if a different disease was diagnosed that 
was potentially risk-modifying for the disease being studied).
If disease $j$ and $k$ share a first step, then the cdf for both $j$ 
and $k$ occurring (in any order), is (see Appendix \ref{rates}), 
\begin{equation}
F_{jk}(t)=\lambda_0\lambda_j\lambda_k c_{jk}(q_0,q_k,q_k) t^{q_0+q_j+q_k+3}
\end{equation}
that can also be estimated from observational data. 
Therefore the incident rates conditional on having already observed one 
disease can be estimated from,   
\begin{equation}\label{fitEq1}
F_{k|j}(t)=\frac{F_{jk}(t)}{F_j(t)} = \lambda_k t^{q_k+1} \kappa_1(q_0,q_j,q_k)
\end{equation}
and, 
\begin{equation}
F_{j|k}(t)=\frac{F_{jk}(t)}{F_k(t)} = \lambda_j t^{q_j+1} \kappa_2(q_0,q_j,q_k)
\end{equation}
that solely involves $q_k$ and $q_j$ respectively, and the terms 
$\kappa_1$, and $\kappa_2$ are functions of the 
powers $q_0,q_j,q_k$, with $\kappa_1$ and $\kappa_2$ having the same form but 
with $j$ and $k$ swapped. 
In contrast, the ratio, 
\begin{equation}\label{fitEq3}
\frac{F_{j}(t) F_k(t)}{F_{jk}(t)} 
= \lambda_0 t^{q_0+1} \kappa_3(q_0,q_j,q_k)
\end{equation}
only involves $q_0$. 
As a simple estimate and proof-of-principle, 
equations \ref{fitEq1}-\ref{fitEq3} allow the powers $q_0,q_j,q_k$ to be determined  
from a log-log plot (as in figure \ref{hazFig}), possibly with 
appropriate left-truncation as described in \cite{Webster2022}. 
In principle the powers $q_0,q_j,q_k$ can also allow the constants 
$\kappa_1$, $\kappa_2$, and $\kappa_3$, to be evaluated,  
and subsequent determination of the rate constants $\lambda_0$, $\lambda_j$, and 
$\lambda_k$. 
Therefore the parameters, including those for the unobserved 
shared step, can all in principle be determined from the data. 

\begin{figure}
	\centering 
	\includegraphics[width=0.8\linewidth]{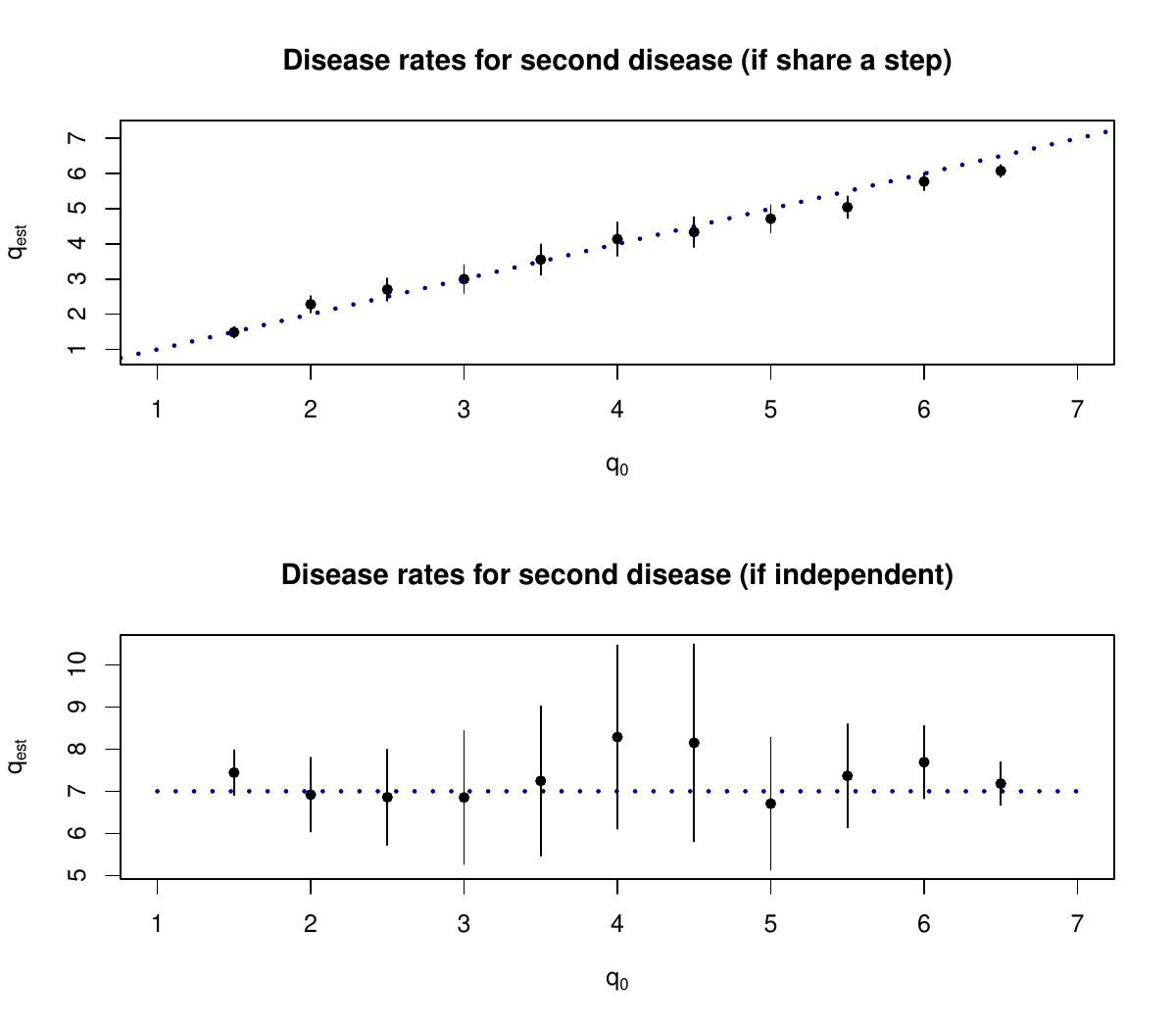}
	\caption{Simulated data illustrate the difference between 
		incidence rates when two diseases are independent, and when 
		they share a first step, changing $q_0$ but keeping $q_0 + q_j+1=7$. 
		If only using data after the first disease has occurred, then 
		if the diseases are independent then their incidence rates will 
		equal disease rates without prior disease (bottom figure, dotted line), 
		but if diseases share a first step then the next disease's incidence 
		rate will depend on 
		the rate of the first step (top figure, dashed line). 
	}\label{fig2}
\end{figure}

In practice the data can be fit using a maximum likelihood method 
for example, such as that described in \cite{Webster2022} and 
illustrated in figure \ref{hazFig}. 
Figure \ref{fig2} illustrates the application of the method 
to simulated data, for rates of disease $j$ after disease $k$ has been 
observed and with $q_0+q_j=6$. 
For diseases with a shared step and $F_0(t)=\lambda_0 t^{q_0+1}$, then if disease 
$k$ has occurred then the 
pdf for disease $j$ that is fit by the maximum likelihood method is 
$d/dt(F_{j|k})\propto t^{q_j} = t^{6-q_0}$, in agreement 
with figure \ref{fig2} where $q_{est}=6-q_0$. 
For independent diseases, the pdf that is fit by the maximum likelihood method 
is $d/dt(F_j) \propto t^{q_0+q_j+1} = t^7$, which is again consistent 
with figure \ref{fig2}, with $q_{est}=7$. 
Note that there are much greater error bars for the second case 
because the risk of two diseases is much lower if they are independent 
of each other, leading to fewer cases. 
Further development can allow the rate parameters to be adjusted for 
potential risk factors, and could allow an exposure such as smoking to 
have different associations with the rates of the first 
and second steps. 
For example, it is possible that smoking does not substantially modify 
cholesterol levels, but does modify the processes involved in 
blood-clot related disease risks. 
In that case, stopping smoking before the latter disease stages where plaque formation and instability are a risk, would be expected to substantially 
reduce the increases in (blood-clot related) disease risks that would 
usually be associated with smoking. 
If this hypothetical example were true, then targeted medication 
in later life might also be able to reduce smoking-related increases 
in clot risk. 
More generally, the approach has the potential to allow large 
observational studies of disease incidence to provide quantitative 
insights into the influence of exposures on biological processes. 



\section{The mathematical model}\label{Model}

This section briefly summarises the mathematical properties of the 
shared step model, and contrasts it with those of 
independent diseases. 
If disease risks are independent of each other then it has been shown that 
the number of diseases that you would expect an individual to have experienced 
by age $t$ is \cite{Webster2021}, 
\begin{equation}
E_{indp}[N]= \sum_j F_j(t)
\end{equation} 
with, 
\begin{equation}
Var_{indp}[N]= \sum_j F_j(t)(1-F_j(t))
\end{equation} 
where $F_j(t)$ is the cumulative distribution function (cdf) for the $j$th disease. 
The probability of surviving without experiencing any disease $j=1..m$ is, 
\begin{equation}\label{pindp0}
P_{indp}(N=0)=P_{indp}(x_1=0, ... , x_m=0) = \Pi_j S_j(t)
\end{equation}
where the age-dependence is implicit on the left side of the equation, 
and $S_j(t)$ is the survival function for disease $j$. 
The Appendix derives equivalent results for diseases that share a step as in 
figure \ref{DAG}. 
The expected number of diseases are unchanged, with,  
\begin{equation}
E[N]=E_{indp}[N]
\end{equation} 
and, 
\begin{equation}
Var[N]= Var_{indp}[N] 
+ 
\left( 
\sum_{j\neq k}  
\int_0^t f_0(t_0)  F_{0j} F_{0k} dt_0 \right)
- \sum_{j\neq k}  
\left( \int_0^t f_0(t_0) F_{0j} dt_0 \right) 
\left( \int_0^t f_0(t_0) F_{0k} dt_0 \right) 
\end{equation} 
where $f_0(s)$ is the pdf for the first stage of the disease process, 
and $F_{0j}(t)$ is the cdf for the second stage of disease $j$. 
With those definitions, the cdf for disease $j$ is \cite{Webster2021}, 
\begin{equation}\label{Fj0}
F_j(t)= \int_0^t f_0(s) F_{0j}(t-s) ds 
\end{equation}
The generating function for the moments is now, 
\begin{equation}
G(s)= \left( 1 - F_0(t) \right) + 
\int_0^t f_0(t_0) \Pi_j \left( 1 - F_{0j}(t-t_0) + e^s F_{0j}(t-t_0)  \right) 
\end{equation}
The probability of surviving without experiencing any disease $j=1..m$ is modified 
from Eq. \ref{pindp0} to, 
\begin{equation}\label{P0}
P(N=0)=P(x_1=0, ... , x_m=0) = S_0(t) + \int_0^t F_0(s) \Pi_j S_{0j}(t-s) ds 
\end{equation}
where $S_{0}(t)$ and $F_0(t)$ are the survival function and cdf for the first step, and 
$S_{0j}(t)$ and $F_{0j}(t)$ are the survival function and cdf for the subsequent step 
that is needed for disease $j$. 
Because, 
\begin{equation}\label{ENpositive}
E[N|N>0] =\frac{E[N]}{1-P(N=0)} 
\end{equation}
then given the observation of at least one disease, then the total number of 
diseases that you are expected to observe is greater if they are not independent, 
if and only if  $P(N=0)>P_{indp}(N=0)$.




\section{Example 1: Genetic mutations with constant hazard rates}\label{examples}

Consider processes with constant hazard rates, as might be a 
reasonable model for genetic mutations. 
Take the first step to have hazard $\lambda_0$, and subsequent steps $1...m$ to 
have hazards $\lambda_j$, so $f_0(t)=\lambda_0 e^{-\lambda_0 t}$ and 
$S_j(t)=e^{-\lambda_j t}$. 
Now change variables so that $\lambda_0 t$ becomes $t$ and $\lambda_j/\lambda_0$ 
becomes $\lambda_j$, leaving 
$f_0(t)=e^{-t}$ and $S_j=e^{-\lambda_j t}$. 
If we had $m$ {\sl independent} diseases whose first step occurs with the same
rate $\lambda_0$, but are otherwise independent with different rates $\lambda_j$ 
for the second step, then Eq. \ref{Fj0} gives the 
cumulative distribution function $F_j(t)=1-S_j(t)$ for disease $j$ as,
\begin{equation}
\begin{array}{ll}
F_j(t) 
&=\int_0^t f_0(s) \left( 1 - S_{1j}(t-s) \right) ds 
\\
&=\int_0^t e^{-s} \left( 1 - e^{-\lambda_j(t-s)} \right) ds 
\\
&=\left( 1- e^{-t} \right) - 
\frac{e^{-\lambda_j t}}{(1-\lambda_j)} 
\left( 
1 - e^{-(1-\lambda_j)t}
\right)
\\
&= 1 - \left( \frac{ e^{-\lambda_jt} - \lambda_j e^{-t} }{ 1-\lambda_j } \right) 
\end{array}
\end{equation} 
Eq. \ref{pindp0} then gives the probability of zero diseases being 
observed by age $t$ as,  
\begin{equation}
\begin{array}{ll}
P_{indp}(0) 
&= \Pi_j \left( 1 - F_j(t) \right)
\\
&= \Pi_j \left( \frac{ e^{-\lambda_j t} - \lambda_j e^{-t} }{1-\lambda_j} \right) 
\end{array}
\end{equation}
In contrast, if $m$ diseases have a shared first step, then 
using Eq. \ref{P0},  
\begin{equation}
\begin{array}{ll}
P(0) 
&= \left( 1 - F_0(t) \right) 
+ \int_0^t f_0(s) \Pi_j S_{0j}(t-s) 
\\
&= e^{-t} + \int_0^t e^{-s} \Pi_j e^{-\lambda_j (t-s)}
\\
&= e^{-t} + \frac{e^{-m\bar{\lambda}t}}{1-m\bar{\lambda}} 
\left( 1 - e^{-(1-m\bar{\lambda})t}   \right) 
\\
&= \frac{e^{-m\bar{\lambda}t} - m\bar{\lambda} e^{-t} }{1-m\bar{\lambda}} 		
\end{array}
\end{equation}
where $\bar{\lambda}=\frac{1}{m} \sum_{j=1}^m \lambda_j$. 
By firstly showing that $P(0)>P_{indp}(0)$, we can use Eq. \ref{ENpositive} to show that 
$E[N|N>0]$ is greater for diseases with a shared step, 
than it would be if diseases were independent. 
To see this, firstly note that we can write $P(0)$ as, 
\begin{equation}\label{P01}
\begin{array}{ll}
P(0) 
&= \frac{e^{-m\bar{\lambda}t} - m\bar{\lambda} e^{-t}}{1-m\bar{\lambda}} 
\\ 
&= e^{-t} +  
\frac{e^{-m\bar{\lambda}t} - e^{-t}}{1-m\bar{\lambda}} 
\end{array}
\end{equation}
or as,
\begin{equation}\label{P02}
\begin{array}{ll}
P(0) 
&= \frac{e^{-m\bar{\lambda}t} - m\bar{\lambda} e^{-t}}{1-m\bar{\lambda}} 
\\ 
&= e^{-m\bar{\lambda}t} +  \frac{m\bar{\lambda}}{1-m\bar{\lambda}} 
\left( e^{-m\bar{\lambda}t} - e^{-t} \right)
\end{array}
\end{equation}
Then by examining cases with $m\bar{\lambda}>1$ and $0<m\bar{\lambda}<1$, indicates 
that the right-hand term in both equations are positive, and hence that 
$P(0)>e^{-t}$ and $P(0)> e^{-m\bar{\lambda}t}$ for all $0<m\bar{\lambda}$. 
Next write, 
\begin{equation}
\begin{array}{ll}
P_{indp}(0) 
&= \Pi_{j=1}^m 
\frac{e^{- \lambda_j t} - \lambda_j e^{-t}}{1-\lambda_j} 
\\ 
&= \exp m \left\{ \frac{1}{m} \sum_{j=1}^m 
\log \left( \frac{e^{- \lambda_j t} - \lambda_j e^{-t}}{1-\lambda_j} \right) \right\} 
\end{array}
\end{equation}
and use Jensen's inequality to infer that,  
\begin{equation}
\frac{1}{m} \sum_{j=1}^m \log f(\lambda_j) \leq \log f \left( \frac{1}{m} \sum_{j=1}^m \lambda_j \right) 
\end{equation}
and as a consequence, that, 
\begin{equation}
\begin{array}{ll}
P_{indp}(0) 
& \leq \exp m \left\{  \log \left(  \frac{e^{-\bar{\lambda}t} - \bar{\lambda} e^{-t}}{1-\bar{\lambda}} \right) \right\}
\\
& = \left(  \frac{e^{-\bar{\lambda}t} - \bar{\lambda} e^{-t}}{1-\bar{\lambda}} \right)^m
\end{array}
\end{equation}
Then considering $(e^{-\bar{\lambda}t} - \bar{\lambda} e^{-t})/(1-\bar{\lambda})$, in 
a similar way to Eq. \ref{P01} and \ref{P02}, and   
examining cases $\bar{\lambda}>1$ and $\bar{\lambda}<1$, then it can be seen 
that either $P_{indp}(0)< e^{-mt}$ or $P_{indp}(0)<e^{-m\bar{\lambda}t}$. 
Therefore when $(e^{-\bar{\lambda}t} - \bar{\lambda} e^{-t})/(1-\bar{\lambda})$ is 
raised to the power $m$, then using Eq.  \ref{P01} and \ref{P02} we have for each case,  
\begin{equation}
P_{indp}(0) < e^{-mt} < e^{-t} < P(0)
\end{equation}
and,
\begin{equation}
P_{indp}(0) < e^{-m\bar{\lambda}t} < P(0)
\end{equation}
For $P_{indp}(0)< P(0)$, then $1-P(0) < 1- P_{indp}(0)$, so noting that 
$E[N]$ is the same for both independent diseases and for diseases with a shared step 
(Eq. \ref{EN}), then, 
\begin{equation}
\begin{array}{ll}
E[N|N>0] &=
\frac{E[N]}{1-P(0)}
\\
&>\frac{E[N]}{1-P_{indp}(0)}
\\
&=E_{indp}[N|N>0]
\end{array}
\end{equation}
Therefore, although $E[N]=E_{indp}[N]$ (Eq. \ref{EN}), because $P(0)>P_{indp}(0)$ 
then for this example $E[N|N>0] > E_{indp}[N|N>0]$. 
Because of this and irrespective of risk factors, for this scenario with constant 
hazard rates and several diseases sharing a first step, we would expect to see 
some people with more diseases than expected (compared with independent diseases), 
and more people than expected with none. 

\section{Discussion}\label{Discussion}

In the 1950s, the multistage model of carcinogensis led  Prof. Peter Armitage 
and Sir Prof. Richard Doll to perform one of the first large-scale population 
studies for the risk of individual cancer types. 
It is remarkable that over 75 years later, in the year of 
Prof. Peter Armitage's death, that it has suggested a valuable conceptual model 
for studying the risk of two or more diseases.  
The model can be improved in several ways, such as allowing adjustment for 
risk factors - that can be different for the different stages in the 
disease process, and for different diseases. 
Then large observational studies of disease 
incidence would have the potential to provide quantitative 
insights into the influence of external exposures or 
genetic variation on biological processes. 
The model can potentially be reformulated as a non-parametric 
model, possibly similar to the conventional Cox proportional hazards model 
but accounting for a shared step and different disease risks. 
More than one shared step, or a shared step at later disease stages are 
obvious variants of the model, although the conceptually simple model 
with only two steps may provide greater insights in the short term. 
Although conceived in the context of multimorbidity, the model 
can also be used to model diseases that require a similar set of genetic 
mutations prior to disease initiation, this being the disease type that 
the multistage model was originally intended to describe.  

\appendix

\section*{Appendix}

\section{Properties of the model: Generating function, expectation and variance}\label{gf}


The key to the mathematical derivation is the observation that 
although the onset rates of diseases are not independent, they 
are conditionally independent 
after the first shared step has occurred. 
Consider the probability of observing $N$ of $m$ possible diseases with the same shared 
step ``$0$'', occurring within time $t$, 
\begin{equation}
P(N)=\sum_{\{x_j=0,1\}} \delta(N-\sum_j x_j) P( X_1=x_1, ... , X_m=x_m )
\end{equation}
where $P( X_1=x_1, ... , X_m=x_m )$ is the probability of observing the set of diseases $1..m$, 
and $X_j=1$ if disease $j$ has occurred and $X_j=0$ otherwise, and 
$\delta(N-\sum_j x_j)$ equals $1$ if $N=\sum_j x_j$ but is otherwise zero 
(similarly to Eq. 1 of \cite{Webster2023}).  
This can be written as, 
\begin{equation}
P( X_1=x_1, ... , X_m=x_m )= P( X_1=x_1, ... , X_m=x_m, t_0<t ) + P( X_1=x_1, ... , X_m=x_m, t_0>t )
\end{equation}
where the conditions $t_0 <t$ and $t_0>t$ indicate whether disease onset has occurred before 
or after the time $t_0$ at which the first step occurred. 
If $t_0>t$ then the diseases cannot have occurred because the first step $0$ is necessary 
prior to disease onset, and as a result,  
\begin{equation}\label{tless}
\begin{array}{ll}
P( X_1=x_1, ... , X_m=x_m, t_0>t ) 
& = P( X_1=x_1, ... , X_m=x_m | t_0>t ) P(t_0>t)
\\
& = \delta( 0=x_1, ... , 0=x_m ) P(t_0>t)
\end{array}
\end{equation}
In terms of pdf $f(t)$ and cdf $F(t)$, then $P(t_0>t)=(1-F_0(t))$, where $F_0(t)$ is 
the cdf for the first step. 
Once the first step has occurred, then the subsequent risk of onset of 
diseases $1, ... , m$ are assumed to be independent, and we have, 
\begin{equation}\label{tmore}
\begin{array}{ll}
P( X_1=x_1, ... , X_m=x_m, t_0<t ) 
&= \int_0^t P( X_1=x_1, ... , X_m=x_m, T_0=t_0 ) dt_0
\\
&=  \int_0^t P( X_1=x_1, ... , X_m=x_m | T_0=t_0 ) f_0(t_0) dt_0
\\
&=  \int_0^t f_0(t_0) 
\Pi_{j=1}^m F_{0j}(t-t_0)^{x_j} ( 1- F_{0j}(t-t_0) )^{1-x_j} dt_0
\end{array}	
\end{equation}
where $f_0(t_0)$ is the pdf for the first step occuring at time $t_0$, and 
$F_{0j}(t-t_0)^{x_j} ( 1- F_{0j}(t-t_0) )^{1-x_j}$ is the probability of 
disease $j$ occuring within the remaining time $t-t_0$ if $x_j=1$, or not if 
$x_j=0$. 
Therefore using Eqs \ref{tless}, \ref{tmore}, and $P(t_0>t)=(1-F_0(t))$,
\begin{equation}\label{cdf}
\begin{array}{ll}
P( X_1=x_1, ... , X_m=x_m) 
&= (1-F_0(t)) \delta( 0=x_1, ... , 0=x_m )	
\\
&+ \int_0^t f_0(t_0) 
\Pi_{j=1}^m F_{0j}(t-t_0)^{x_j} ( 1- F_{0j}(t-t_0) )^{1-x_j} dt_0
\end{array}
\end{equation}
The generating function is, 
\begin{equation}\label{Gs}
\begin{array}{ll}
G(s)
&=\sum_{N=0}^m e^{sN} P(N)
\\
&=\sum_{N=0}^m e^{sN} \sum_{\{x_j=0,1\}} \delta(N-\sum_j x_j) P( X_1=x_1, ... , X_m=x_m )
\\
&=\sum_{\{x_j=0,1\}} e^{s\sum_j x_j} P( X_1=x_1, ... , X_m=x_m )
\\
&=\sum_{\{x_j=0,1\}} \left( \Pi_j e^{sx_j} \right)
\left\{   
\left( 1 - F_0(t) \right) \Pi_j \delta(0=x_j) 
+ \int_0^t f_0(t_0) \Pi_j F_{0j}(t-t_0)^{x_j} 
\left( 1 - F_{0j}(t-t_0) \right)^{1-x_j} 
\right\}
\\
&=\sum_{\{x_j=0,1\}} 
\left\{
\left( 1 - F_0(t) \right)
\Pi_j \delta(0=x_j) e^{sx_j}
+ \int_0^t f_0(t_0) \Pi_j e^{sx_j} F_{0j}(t-t_0)^{x_j} 
\left( 1 - F_{0j}(t-t_0) \right)^{1-x_j} 
\right\}
\\
&= \left( 1 - F_0(t) \right) + 
\int_0^t f_0(t_0) \Pi_j \left( 1 - F_{0j}(t-t_0) + e^s F_{0j}(t-t_0)  \right) 
\end{array}
\end{equation}
Moments are obtained in the usual way, by differentiating with respect to $s$ and 
then setting $s=0$ in the resulting expression. 
Setting $s=0$ in Eq. \ref{Gs} above gives $G(s=0)=1-F_0(t)+\int_0^t f_0(t_0) dt_0 = 1$, 
as it should do for a normalised probability distribution. 
The first and second derivatives are, 
\begin{equation}\label{first}
\begin{array}{ll}
\frac{\partial G}{\partial s}
&= \int_0^t f_0(t_0) 
\sum_j \frac{e^s F_{0j}(t-t_0)}{ 1-F_{0j}(t-t_0)+e^s F_{0j}(t-t_0) }
\Pi_k \left( 1 - F_{0k}(t-t_0) +e^s F_{0k}(t-t_0) \right) 
\end{array}
\end{equation}
and,
\begin{equation}\label{second}
\begin{array}{ll}
\frac{\partial^2 G}{\partial s^2}
&= \int_0^t f_0(t_0) \sum_j 
\left( 
\frac{e^s F_{0j}(t-t_0)}{1-F_{0j}(t-t_0)+e^sF_{0j}(t-t_0)}  
- \frac{e^se^s F_{0j}(t-t_0)^2}{(1-F_{0j}(t-t_0)+e^sF_{0j}(t-t_0))^2} \right) 
\Pi_k \left( 1 - F_{0k}(t-t_0) +e^s F_{0k}(t-t_0) \right) 
\\
&+ \int_0^t f_0(t_0) 
\left( \sum_j \frac{e^s F_{0j}(t-t_0)}{ 1-F_{0j}(t-t_0)+e^s F_{0j}(t-t_0) }
\right)^2
\Pi_k \left( 1 - F_{0k}(t-t_0) +e^s F_{0k}(t-t_0) \right) 
\end{array}
\end{equation}
Noting that $F_j(t)=\int_0^t f_0(t_0) F_{0j}(t-t_0) dt_0$ and using 
Eqs. \ref{Gs} and \ref{first}, we get, 
\begin{equation}\label{EN}
\begin{array}{ll}
E[N] &= \left. \frac{\partial G}{\partial s} \right|_{s=0} 
\\
&= \int_0^t dt_0 f_0(t_0) \sum_j F_{0j}(t-t_0)
\\
&= \sum_j \int_0^t dt_0 f_0(t_0) F_{0j}(t-t_0)
\\
&= \sum_j F_j(t) 
\end{array}
\end{equation}
which is the same as was previously found for independent diseases 
that do not share a first step \cite{Webster2022}. 
Using Eq. \ref{first} and \ref{second}, then writing $F_j$, $F_{0j}$ 
in place of $F_j(t)$ and $F_{0j}(t-t_0)$ to abbreviate the notation, then, 
\begin{equation}\label{VarN}
\begin{array}{ll}
Var(N) 
&= E[N^2] - E[N]^2 
\\
&=  \left. \frac{\partial G^2}{\partial s^2} \right|_{s=0} 
- \left( \left. \frac{\partial G}{\partial s} \right|_{s=0} \right)^2
\\
&=
\int_0^t f_0(t_0) \left\{ 
\sum_j F_{0j} - \sum_j F_{0j}^2 + \sum_{j,k} F_{0j}F_{0k}
\right\}dt_0 
-   \left( \int_0^t f_0(t_0) \sum_j F_{0j} dt_0 \right)^2 
\\
&=
\sum_j F_{j} \left( 1 - F_j \right) 
+ \left(
\sum_{j\neq k}  
\int_0^t f_0(t_0)  F_{0j} F_{0k} dt_0 \right)
- \sum_{j\neq k}  
\left( \int_0^t f_0(t_0) F_{0j} dt_0 \right) 
\left( \int_0^t f_0(t_0) F_{0k} dt_0 \right) 
\end{array}
\end{equation} 
where Eq. \ref{EN} gave $E[N]=\sum_j F_j =  \int_0^t f_0(t_0) \sum_j F_{0j}$, and 
the first and last terms in the 2nd line from bottom can be grouped 
to give $\sum_j F_{j} \left( 1 - F_j \right)=Var_{indp}(N)$, which is 
the variance if the diseases were independent \cite{Webster2022} and not sharing 
a step. 
The final line of Eq. \ref{VarN} can be re-expressed in several different ways, 
such as using Eq. \ref{p2} for the probability of observing both diseases $j$ and $k$ 
if they share a step ($F_{jk}$), that for most realistic scenarios will be much 
smaller than $F_jF_k$ (the probability of observing $j$ and $k$ if they were 
equivalent risk for either $j$ or $k$ alone, but were independent of each other). 
Therefore in most situations we expect $Var(N) > Var_{indp}(N)$, but this 
does not appear to be guaranteed for all times $t$. 

\section{Expected number of diseases}\label{EN-sec}

Eq. \ref{EN} found that if diseases share a first step, then the expected 
number of disease cases is the same as for diseases that are independent 
(without a shared first step) \cite{Webster2022}.
This at first seems surprising.  
When there is a shared step, we might expect disease incidence to 
be more common after at least one disease (with a shared first step), has occurred. 
However, everyone will always have at least one disease (if they are to die). 
Therefore, consider instead, $E[N|N>0]$, 
\begin{equation}\label{ENg1}
\begin{array}{rl}
E[N|N>0] 
&= \sum_{N=0}^{\infty} N P(N|N>0) 
\\
&= \sum_{N=0}^{\infty} N \frac{P(N,N>0)}{P(N>0)} 
\\
&= \sum_{N=0}^{\infty} N \frac{P(N)}{P(N>0)} \text{  because summing over N=0}
\\
&=\frac{E[N]}{P(N>0)}
\\
&=\frac{E[N]}{1-P(N=0)} 
\end{array}
\end{equation}
To evaluate this we need to be able to calculate $P(N=0)$. 
For diseases with a shared first step, then Eq. \ref{cdf} gives, 
\begin{equation}
\begin{array}{rl}
P( 0=x_1, ... , 0=x_m) 
&= 1-F_0(t) 
+ \int_0^t f_0(s) \Pi_j ( 1- F_{0j}(t-s) ) ds
\end{array} 
\end{equation}
For independent diseases, 
\begin{equation}
\begin{array}{rl}
P_{indp}(0=x_1, ... 0=x_m) 
&=\left. \Pi_j F_j(t)^{x_j}(1-F_j(t))^{1-x_j} \right|_{x_j=0}
\\
&=\Pi_j  \left( 1-F_j(t) \right)
\end{array} 
\end{equation}
That may alternately be written in terms of survival functions, e.g.  
$S_0(t)=1-F_0(t)$, with 
\begin{equation}\label{P0app}
P( 0=x_1, ... , 0=x_m) 
= S_0(t) 
+ \int_0^t f_0(s) \Pi_j S_{0j}(t-s)  ds
\end{equation}
and, 
\begin{equation}\label{P0indp}
P_{indp}(0=x_1, ... 0=x_m) 
=\Pi_j  \left( 1-F_j(t) \right) = \Pi_j  S_j(t) 
\end{equation}
Eq. \ref{P0indp} shows that all diseases must be avoided 
(``survived''), if no disease is to be observed.  
Eq. \ref{P0app} can be interpreted as disease being avoided by either 
surviving without the first step occurring ($S_0(t)$), or 
by the first step having occurred, but surviving all of the 
subsequent steps needed to trigger disease 
$\int_0^t f_0(s) \Pi_j S_{0j}(t-s)ds$.

\section{Incidence rates for multistage diseases with a shared step}\label{rates}

Using Eq. \ref{p2}, the probability of observing both diseases $j$ and $k$  by age $t$, is, 
\begin{equation}
F_{jk}(t) = \int_0^t f_0(s) F_{0j}(t-s) F_{0k}(t-s) ds 
\end{equation}
Assuming a multistage model with for disease processes that occur with 
cdfs that can be approximated by: 
\begin{equation}
\begin{array}{c}
F_0(t) = \lambda_0 t^{q_0+1} \\
F_{0j}(t)=\lambda_j t^{q_j +1} \\
F_{0k}(t)=\lambda_k t^{q_k +1} 
\end{array}
\end{equation}
with pdf for step $0$ of $f_0(t) = (q_0+1) \lambda_0 t^{q_0}$. 
Then Eq. \ref{p2} gives, 
\begin{equation}\label{fija}
F_{jk}(t) = \int_0^t (q_0+1) \lambda_0 \lambda_j \lambda_k 
s^{q_0}  (t-s)^{q_j+q_k+2} 
\end{equation} 
Comparison with Eqs. 16 and 21 of reference \cite{Webster2019} indicates that 
Eq. \ref{fija} is mathematically equivalent to a process with two steps, 
the first step with  
$f_0(t) = (q_0+1) \lambda_0 t^{q_0}$ 
and a second step whose pdf  is the derivative of the cdf that appears 
in the convolution given by Eq. \ref{fija}, with pdf, 
\begin{equation} 
\tilde{f}_2(t) = 
\lambda_j 
\lambda_k 
t^{q_j+q_k+2} . 
\end{equation}
Therefore we can use the usual results from Eqs. 16, 21, and 28 of \cite{Webster2022}, 
that evaluate Eq \ref{fija} using Laplace transforms to give, 
\begin{equation}
F_{jk} = 
\lambda_0
\lambda_j 
\lambda_k
c_{jk}(q_0,p_1,p_2) 
t^{q_0+q_j+q_k+3} 
\end{equation}
with, 
\begin{equation}
c_{jk}(q_0,p_1,p_2) =
\frac{(q_0+1)!(q_j+q_k+2)!}{(q_0+p_1+p_2+3)!}
\end{equation}
An equivalent set of calculations give, 
\begin{equation}\label{Fj}
F_j(t)=\lambda_0 \lambda_j 
c_j(q_0,q_j) t^{q_0+q_j+2}
\end{equation}
with, 
\begin{equation}\label{cj}
c_j(q_0,q_j) 
= \frac{(q_0+1)!(q_j+1)!}{(q_0+q_j+2)!} 
\end{equation}
$F_k$ is obtained by replacing $\lambda_j$ with $\lambda_k$ in 
Eqs. \ref{Fj}, \ref{cj}, above, with $c_k(q_0,q_k)=c_j(q_0,q_k)$. 
The terms $\kappa_j$, $\kappa_k$ and $\kappa_{jk}$ are,
\begin{equation}
\kappa_j = \frac{c_{jk}}{c_j} 
= \frac{(q_j+q_k+2)!(q_0+q_j+2)!}{(q_0+q_j+q_k+3)!(q_j+1)!} ,
\end{equation}
\begin{equation}
\kappa_k = \frac{c_{jk}}{c_k} 
= \frac{(q_j+q_k+2)!(q_0+q_k+2)!}{(q_0+q_j+q_k+3)!(q_k+1)!} ,
\end{equation}
and, 
\begin{equation}
\kappa_{jk} = \frac{c_j c_k}{c_{jk}} 
= \frac{(q_0+q_j+q_k+3)!(q_j+1)!(q_k+1)!(q_0+1)!}{(q_j+q_k+2)!(q_0+q_j+2)!(q_0+q_k+2)!}.
\end{equation}

\section{Example 2: Two diseases with a shared step}\label{ex2}

Multistage models are commonly associated with ``power-law'' rates 
of incidence $\sim t^p$ for an age $t$ and a power $p$ that was 
originally conceived as being an integer \cite{AD1954}, but need 
not be so in general \cite{Webster2019,Webster2022}.  
This example is intended to explain how the power-law incident rates of 
multistage models arise from the results described here, for the usual 
situation of rare diseases (most diseases are rare, in 
the sense that we are very unlikely to experience them, it is the very 
large number of diseases that makes disease inevitable with increasing 
age \cite{Webster2022}).   
If considering only two disease types $j$ and $k$ for example, then the 
probability distribution can be obtained from Eq. \ref{cdf} either 
by setting $m=2$ or by marginalising over $x_r$ for $r\neq j,k$. 
This gives the probability of observing the two diseases 
within a time $T<t$ (that implicitly requires $t_0<t$), as,   
\begin{equation}
\begin{array}{ll}
P( X_j=x_j, X_k=x_k ) 
&= (1-F_0(t)) \delta( 0=x_j, 0=x_k )
\\	
&+ \int_0^t f_0(t_0) 
F_{0j}(t-t_0)^{x_j} F_{0k}(t-t_0)^{x_k} 
( 1- F_{0j}(t-t_0) )^{1-x_j} dt_0 ( 1- F_{0k}(t-t_0) )^{1-x_k} dt_0
\end{array}
\end{equation}
If we are considering the probability of both diseases occurring (with 
$X_j=1$ and $X_k=1$), as opposed to the probability of an outcome 
$(X_j, X_k)$ that may have either or both of $X_j$ and $X_k$ equal to zero, 
then, 
\begin{equation}\label{p2}
\begin{array}{ll}
P( X_j=1, X_k=1 ) 
&= \int_0^t f_0(t_0) 
F_{0j}(t-t_0) F_{0k}(t-t_0) dt_0
\end{array}
\end{equation}
that we might intuitively have written down without derivation. 
We can proceed similarly for 3 or more diseases. 
Considering two diseases, then the probability of observing disease $j$ 
if disease $k$ has already occurred, is, 
\begin{equation}
P( X_j=1 | X_k=1 )
= \frac{\int_0^t f_0(t_0) F_{0j}(t-t_0)  F_{0k}(t-t_0)}{\int_0^t f_0(t_0) F_{0k}(t-t_0)}
\end{equation}
If $f_0(t)$ is normalised by a factor $1/F_0(t)$, then it becomes a pdf over the 
interval $[0,t]$. 
Denoting expectations with respect to this pdf as $E_0[]$, then we can write the 
above equation as, 
\begin{equation}\label{E30}
\frac{P( X_j=1 | X_k=1 )}{P(X_j=1)}
= \frac{F_0(t) E_0[F_{0j}(t-t_0)F_{0k}(t-t_0)]}{F_0(t)^2 E_0[F_{0j}(t-t_0)]E_0[F_{0k}(t-t_0)]}
\propto \frac{1}{F_0(t)}
\end{equation}
This is consistent with the observation that rates are increased in individuals 
that have had one or more disease \cite{Webster2022}, but the time-dependent 
behaviour depends on the correlation between $F_{0j}$ and $F_{0k}$. 
For example, if disease $j$ and $k$ had the same cdfs, with $F_{0j}=F_{0k}$, then 
using Jensen's inequality we would have 
$P( X_j=1 | X_k=1 )/P(X_j=1)\geq (1/F_0(t))$, 
but more generally the result will depend on the specific forms of distribution. 
This has the potential to explain the observed increase in disease risk after people 
have had at least one disease \cite{Webster2022}, and why disease risk was found to 
be approximately proportional to the underlying risk in disease-free 
individuals \cite{Webster2022}. 
It was this latter observation that originally motivated this work. 
However, the following example shows that in general it may not be so simple. 
Returning to the example from Section \ref{examples} with constant rates for each step, 
then, 
\begin{equation}\label{f12} 
\begin{array}{ll}
P( X_j=1 | X_k=1 )
&= (1-e^{-t})  
- \frac{e ^{-t} - e^{-\lambda_j t}}{\lambda_j -1}
- \frac{e^{-t}-e^{-\lambda_k t}}{\lambda_k -1}
+\frac{e^{-t}-e^{-(\lambda_j +\lambda_k)t}}{\lambda_j + \lambda_k -1}
\\
&= F_{\lambda_j} + F_{\lambda_k} - F_{\lambda_j+\lambda_k}
\end{array}
\end{equation}
with $F_{\lambda_j}\equiv \int_0^t \lambda_0 e^{-\lambda_0 s} ( 1 - e^{-\lambda_j(t-s)}) ds$ 
and units changed with $\lambda_0 s$ becoming $s$ and $\lambda_j/\lambda_0$ 
becoming $\lambda_j$ as in Section \ref{examples}. 
Expanding Eq. \ref{f12} in powers of $t$, the first three terms cancel to leave, 
\begin{equation}\label{pjgk}
P( X_j=1 | X_k=1 )
= \lambda_0 \lambda_j \lambda_k \frac{t^3}{3} + O (t^4)
\end{equation}
a result that we would expect 
to find for a 3-stage process (step 0, disease j or k, then the second disease). 
After integration, 
\begin{equation}
P(X_j=1) = F_{\lambda_j} = (1-e^{-t}) + \frac{ e^{-t} - e^{-\lambda_j t} }{1-\lambda_j}
\end{equation}
Then expanding in $t$ and converting back to original units ($t$ becomes $\lambda_0 t$ and 
$\lambda_j$ becomes $\lambda_j/\lambda_0$), 
to give the quadratic approximation, 
\begin{equation}\label{pj}
P(X_j=1) = \lambda_0\lambda_j \frac{t^2}{2} 
\end{equation}
as expected for a 2-step process.
Eq. \ref{E30} can be evaluated for this example, finding,
\begin{equation}\label{ratiojkj}
\frac{P( X_j=1 | X_k=1 )}{P(X_j=1)}
= \lambda_k \frac{2t}{3} 
\end{equation}
Provided the approximation remains valid, then 
for $\lambda_k t <3/2$ then $P( X_j=1 | X_k=1 )< P(X_j=1)$, 
but if $\lambda_k t > 3/2$ then $P( X_j=1 | X_k=1 )> P(X_j=1)$. 
The example emphasises 
that the behaviour can vary, and 
suggests that we can have $P( X_j=1 | X_k=1 )< P(X_j=1)$ for 
short times (younger ages), 
but $P( X_j=1 | X_k=1 )> P(X_j=1)$ at longer times (older ages). 
If $\lambda_j$ is different to $\lambda_k$, then the incidence rates of the 
second disease could be very different, depending on whether disease $j$ 
has occurred first or disease $k$. 
Comparison of Eqs. \ref{pjgk}, \ref{pj}, and \ref{ratiojkj}, again 
indicate that if there is a shared step, then different 
(age-dependent) rates of disease incidence will be observed 
after a disease that shares the first step has been observed. 

\section*{Data availability}

No new data were used in this study. 
Simulated data were used to produce the figures - see code availability.  

\section*{Code availability}


R code used to produce figures from simulated data will become available 
on publication of the final article. 
\noindent
R packages used in this study include survival\cite{survival}, 
grr\cite{grr}, data.table\cite{data.table}, and maxLik\cite{maxLik2011}. 

\section*{Funding}
No funding was received for this work. 

\section*{Competing interests}
The author declares no competing interests. 

\bibliographystyle{naturemag}
\bibliography{MS2-refs,MultistageReferences-Converted}

\end{document}